\documentclass{ws-ijmpcs}

\usepackage{url,overcite}

\def\trimmarks{}

\def\Dslash{D\!\!\!\!\slash}
\def\nslash{n\!\!\!\slash}
\def\bnslash{\bar n\!\!\!\slash}
\def\pslash{p\!\!\!\slash}

\newcommand{\nn}{\nonumber} 
\newcommand{\bn}{{\bar n}}
\newcommand{\mcdot}{\!\cdot\!}

\newcommand{\be}{\begin{equation}}
\newcommand{\ee}{\end{equation}}

\newcommand{\SCETa}{\mbox{${\rm SCET}_{\rm I}$ }}
\newcommand{\SCETb}{\mbox{${\rm SCET}_{\rm II}$ }}
\newcommand{\SCETg}{\mbox{${\rm SCET}_{\rm G}$ }}

\newcommand{\vect}[1]{\mathbf{#1}}
\newcommand{\abs}[1]{\left\lvert #1\right\rvert}
\newcommand{\bra}[1]{\left\langle #1\right\rvert}
\newcommand{\ket}[1]{\left\lvert #1\right\rangle}
 
\newcommand{\minus}{\!-\!}
\newcommand{\plus}{\!+\!}
\newcommand{\Lqcd}{\Lambda_{\text{QCD}}}

\newcommand{\as}{\alpha_s}

\newcommand{\cO}{\mathcal{O}}

\newcommand{\cI}{\mathcal{I}}
\newcommand{\cL}{\mathcal{L}}

\newcommand{\cB}{\mathcal{B}}
\newcommand{\cP}{\mathcal{P}}
\newcommand{\cE}{\mathcal{E}}
\newcommand{\cD}{\mathcal{D}}

\newcommand{\wt}{\widetilde}

\newcommand{\eq}[1]{Eq.~\eqref{eq:#1}}
\newcommand{\eqs}[2]{Eqs.~\eqref{eq:#1} and \eqref{eq:#2}}

\renewcommand{\sec}[1]{Sec.~\ref{sec:#1}}

\DeclareMathOperator{\Tr}{Tr}
\DeclareMathOperator{\tr}{tr}

\begin{document}

\markboth{C. Lee}
{The Evolution of SCET}

%
\catchline{}{}{}{}{}
%

\title{The Evolution of Soft Collinear Effective Theory}

\author{Christopher Lee}

\address{Theoretical Division, Los Alamos National Laboratory, MS B283 \\
Los Alamos, NM 87544,
USA\\
clee@lanl.gov}

\maketitle

\begin{history}
\received{Day Month Year}
\revised{Day Month Year}
\published{Day Month Year}
\end{history}

\begin{abstract}
Soft Collinear Effective Theory (SCET) is an effective field theory of Quantum Chromodynamics (QCD) for processes where there are energetic, nearly lightlike degrees of freedom interacting with one another via soft radiation. SCET has found many applications in high-energy and nuclear physics, especially in recent years the physics of hadronic jets in $e^+e^-$, lepton-hadron, hadron-hadron, and heavy-ion collisions. SCET can be used to factorize multi-scale cross sections in these processes into single-scale hard, collinear, and soft functions, and to evolve these through the renormalization group to resum large logarithms of ratios of the scales that appear in the QCD perturbative expansion, as well as to study properties of nonperturbative effects. We overview the elementary concepts of SCET and describe how they can be applied in high-energy and nuclear physics.
\keywords{QCD, effective field theory, SCET, factorization, resummation, jets.}
\end{abstract}

\ccode{PACS numbers: 12.38.Cy, 12.39.St, 13.87.-a, 24.85.+p, 25.30.Fj, 25.75.Bh}

\section{Introduction}	

It is fair to say that effective field theory (EFT) has proven to be one of the most powerful tools in modern physics \cite{Weinberg:1978kz,Manohar:1996cq}. By exploiting power expansions in small parameters determined by hierarchies of physical scales, EFTs help us make advances in predictive power in controlled approximations that may be more difficult to implement directly in the context of a full theory. In Quantum Chromodynamics (QCD), the development and application of Soft Collinear Effective Theory (SCET)\cite{Bauer:2000ew,Bauer:2000yr,Bauer:2001ct,Bauer:2001yt,Bauer:2002nz,BauerStewart} in past decade-and-a-half has brought about such advances for physical processes with energetic, nearly light-like degrees of freedom such as jets. SCET has advanced our understanding of $B$ physics and collider and jet physics in vacuum and in heavy-ion collisions. These problems exhibit dependence on the hierarchically separated scales of a hard collision energy $Q$, on a transverse momentum $p_T$ of collinear modes, of soft radiation with momentum $k_s$, and of hadronization at the energy scale $\Lqcd$. SCET facilitates the factorization of physical observables or cross sections dependent on multiple scales into single-scale functions, the resummation of large logarithms of ratios of the scales via renormalization group evolution of these functions, and the demonstration in many cases of the universality of nonperturbative effects on these observables. In some cases, such as hadronic event shapes in $e^+e^-$ collisions, these have led to predictions at the N$^3$LL level of resummed accuracy\cite{Becher:2008cf}, rigorous proof of universality of the leading nonperturbative corrections\cite{Lee:2006fn,Lee:2006nr,Mateu:2012nk}, and consequently highly precise extractions of the strong coupling\cite{Becher:2008cf,Chien:2010kc,Abbate:2010xh}. SCET has been applied widely to jet physics at the Large Hadron Collider (LHC) and in heavy-ion collisions. As jets grow in importance in high-energy and nuclear physics to probe new particles beyond the Standard Model, the quark-gluon plasma, and hadron structure and the strong coupling, at colliders like LHC, the Relativistic Heavy-Ion Collider (RHIC), and a future Electron-Ion Collider (EIC), so also will SCET and its applications. 

In this talk we review the basic elements of SCET. SCET comes in several different flavors, depending on the relevant scales and degrees of freedom in the problem. In this talk we focus mainly on \SCETa for jet cross sections measured with an invariant mass-like variable\cite{Bauer:2000ew,Bauer:2000yr}. There are also \SCETb for jets measured by their internal transverse momentum\cite{Bauer:2002uv,Bauer:2003mga,Chiu:2011qc,Chiu:2012ir} and \SCETg for jets propagating through a dense medium interacting via Glauber modes\cite{Idilbi:2008vm,DEramo:2010ak,Ovanesyan:2011xy}. In \sec{SCET} we will review the construction of these theories. In \sec{factorization} we will illustrate factorization in the context of event shape distributions in DIS, but the ingredients are widely applicable in $ee, pp$ and other processes. In \sec{nonperturbative} we will describe universality of the leading nonperturbative effects in such distributions. Finally in \sec{recent} we will close by just briefly highlighting recent applications and future directions in the field. The talk and proceedings are by nature limited in time and length, and I apologize for all inevitable omissions of important work in this active and expanding field.

\section{Constructing SCET}
\label{sec:SCET}

An EFT is an approximation to another quantum field theory (the ``full theory'') formed by including only the degrees of freedom that are relevant at a given energy scale or in a given kinematic configuration, and ``integrating out'' any additional degrees of freedom. Familiar examples are the effective electroweak theory of the Standard Model, relevant for electroweakly-charged particles with momenta $p\ll M_W$, with the $W,Z$ bosons and top quarks integrated out.\cite{Buras:1998raa} The effective theory Lagrangian can be organized as a series in powers of $p/M_W$, truncated to the desired order of accuracy. A slightly more subtle case is the heavy quark effective theory (HQET)\cite{Eichten:1989zv,Georgi:1990um,Grinstein:1990mj}, in which not the entire heavy quark $Q$ but only its large momentum components are integrated out: $p_Q^\mu = m_Q v^\mu + k^\mu$ where $k^\mu\ll m_Q$ and $v$ is the 4-velocity of an on-shell heavy quark. In all of these cases, matrix elements in the EFT must reproduce those of the full theory in the low-energy or infrared (IR) regime. They differ in the high-energy or ultraviolet (UV) limit. These differences in the UV are encoded in matching coefficients between the full and effective theories.

The power in going through these procedures is that the low-energy EFT may possess additional symmetries or other simplifications to a given order in the power expansion that are not manifest in the full, exact theory. This can increase predictive and computational power. An example is enhanced spin-flavor symmetry in HQET.\cite{Isgur:1989vq} Another, as we shall see below, is soft-collinear decoupling in SCET.\cite{Bauer:2001yt} An EFT is formulated as a power expansion in a small parameter $\lambda$ providing systematic, order-by-order approximations to the full theory. For example, in the effective EW theory, $\lambda = p/M_W$, while in HQET $\lambda= \Lqcd/m_Q$. In SCET, $\lambda = p_T/Q$ where $p_T$ is the transverse momentum of energetic collinear particles within a jet.
We will illustrate here only the construction of the collinear quark part of the Lagrangian in the theory \SCETa\!,\cite{Bauer:2000yr} which follows similar logic as in HQET.\cite{Manohar:2000dt} 

The basic degrees of freedom in SCET are collinear and soft quarks and gluons. Like heavy quarks, collinear degrees of freedom have momenta that can be split into large and small components: $p_c^\mu = \frac{\bn\cdot p}{2}n^\mu + \frac{n\cdot p}{2}\bn^\mu + p_\perp^\mu$,
where $n,\bn = (1,\pm\vect{\hat z})$ are lightlike 4-vectors along the direction $\vect{z}$ in which the collinear particle is traveling, satisfying $n\cdot\bn = 2$, and where $p_\perp$ is orthogonal to both $n,\bn$. Other choices of $n,\bn$ are possible, using reparametrization invariance.\cite{Manohar:2002fd}
For a collinear particle, the lightcone components of its momentum obey a hierarchical scaling, $p^-\equiv \bn\cdot p \sim Q \,, p^+\equiv n\cdot p \sim Q\lambda^2 \,, p_\perp\sim Q\lambda$, where $\lambda\ll 1$. The size of $\lambda$ is determined by the typical invariant mass of the collinear modes being described, $p_c^2 = n\cdot p \,\bn\cdot p + p_\perp^2 \sim Q\lambda^2$, or by the size of $p_\perp$. Soft particles may have momenta $k_s\sim Q\lambda^2$ (\SCETa\!) or $Q\lambda$ (\SCETb\!). The SCET Lagrangian is constructed as a power expansion in $\lambda$, integrating out hard modes of virtuality $p^2\sim Q^2$ from QCD.

We construct here the leading-order collinear quark Lagrangian of \SCETa\!. We begin with the (massless) quark part of the QCD Lagrangian,
\be
\label{eq:DiracLag}
\cL_q = \bar\psi(x)i\Dslash\; \psi(x) \,.
\ee
We factor out of the quark field a large momentum phase factor,
\be
\label{eq:labelphase}
\psi(x) = \sum_{\tilde p \neq 0} e^{-i\tilde p \cdot x} \psi_{n,p}(x)\,,
\ee
where $\tilde p^\mu = \bn\cdot p \frac{n^\mu}{2} + \tilde p_\perp^\mu$ is called a ``label'' momentum, containing the order $Q$ and $Q\lambda$ components of collinear momentum, leaving $\psi_{n,p}$ to describe momentum fluctuations of order $k\sim Q\lambda^2$ about the label momentum. We have a sum over multiple label momenta $\tilde p \neq 0$ since the collinear quarks interact with collinear gluons which can change the large label momentum. The term $\tilde p  =0$ is omitted to avoid overlap and double-counting with soft modes. This is implemented in perturbative calculations via the zero-bin subtraction\cite{Manohar:2006nz}, similar to soft subtractions in full QCD\cite{Lee:2006nr,Berger:2003iw}. Such double-counting must be removed for consistent results. We match the QCD gluon field in the covariant derivative in \eq{DiracLag} onto collinear and soft gluon fields in SCET, $A^\mu \to A_c^\mu + A_s^\mu$, where collinear gluons $A_c$ are given by a similar sum over labels as \eq{labelphase}, and $A_s$ are gluons with soft momenta, $k_s\sim Q\lambda^2$ in \SCETa\!.

We project out large and small components of the collinear Dirac spinors, $\xi_{n,p} = \frac{\nslash\bnslash}{4}\psi_{n,p}$, $\Xi_{n,p} = \frac{\bnslash\nslash}{4}\psi_{n,p}$, in terms of which the Lagrangian \eq{DiracLag} takes the form
\be
\cL_q = \sum_{\tilde p,\tilde p'\neq 0} e^{-i(\tilde p - \tilde p')\cdot x} 
\begin{pmatrix}
\bar\xi_{n,p'} & & \bar\Xi_{n,p'}
\end{pmatrix}
\begin{pmatrix}
\frac{\bnslash}{2} in\cdot D & & \tilde{\smash{\pslash}}_\perp + i\Dslash_\perp \\
 \tilde{\smash{\pslash}}_\perp + i\Dslash_\perp & & \frac{\nslash}{2}(\bn\cdot \tilde p + i\bn\cdot D) \\
\end{pmatrix}
\begin{pmatrix}
\xi_{n,p} \\ \Xi_{n,p}
\end{pmatrix}\,.
\ee
The spinor $\Xi_{n,p}$ acquires a large effective mass, $\bn\cdot\tilde p \sim Q$. So we integrate it out, solving its classical equation of motion order-by-order in $\lambda$ and substituting its solution into $\cL_q$. Some of the explicit label momentum factors showing up in $\cL_q$ can be cleaned up by defining a label momentum operator $\cP^\mu$, acting as $\cP^\mu\phi_{n,p} = \tilde p^\mu \phi_{n,p}$.\cite{Bauer:2001ct}
Then the leading-order collinear quark Lagrangian of SCET can be written
\be
\label{eq:nquarkLag}
\cL_{qn} = \bar \xi_n \Bigl[ in\cdot D + i\Dslash_\perp^{\;c}W_n(x) \frac{1}{n\cdot\cP} W_n^\dag(x) i\Dslash_\perp^{\;c}\Bigr]\frac{\nslash}{2}\xi_n\,,
\ee
where $\xi_n (x) = \sum_{\tilde p } e^{-i\tilde p \cdot x}\xi_{n,p}(x)$, $D_\mu^c = \cP^\mu -igA_\mu^c$, and $\cD^\mu = \bn\cdot \cP \frac{n^\mu}{2} + \cP_\perp^\mu + in\cdot D\frac{\bn^\mu}{2}$.
The Wilson line $W_n = P\exp\bigl[ig\int_{-\infty}^x ds\,\bn\mcdot A^c(\bn s)\bigr]$ is a path-ordered exponential of collinear gluons, required by collinear gauge invariance and representing gluons collinear to $n$ emitted by energetic particles in other directions.\cite{Bauer:2000yr,Bauer:2001ct}

The derivation of the collinear gluon Lagrangian is similar.\cite{Bauer:2001yt} Soft modes by themselves simply obey the Lagrangian of full QCD, $\cL_s = \cL_{\text{QCD}}[q_s,A_s]$. At leading order in $\lambda$ there are no couplings of soft quarks to collinear modes. The SCET Lagrangian up to second subleading order is known.\cite{Bauer:2003mga,Beneke:2002ph,Chay:2002vy,Pirjol:2002km} Alternative formulations of SCET exist in position space without labels\cite{Beneke:2002ph,Becher:2014oda} and in terms of QCD fields\cite{Bauer:2008qu,Freedman:2011kj,Feige:2013zla}.

Now, the only place that soft gluons appear in the Lagrangian \eq{nquarkLag} is in the light-cone component of the covariant derivative $n\cdot D$. This means the coupling of soft gluons to collinear quarks is only through the vector $n^\mu$ rather the full Dirac matrix $\gamma^\mu$. This is a manifestation of the eikonal approximation, made explicit at leading order in $\lambda$ in the SCET Lagrangian. We can take further advantage of this property by making a field redefinition of the collinear fields,
\be
\label{eq:BPS}
\xi_n(x) = Y_n(x) \xi_n^{(0)}(x)\,,
\ee
where $Y_n$ is a Wilson line of soft gluons, $Y_n(x) = P\exp\bigl[ig \int_{-\infty}^x ds\,n\cdot A_s(ns + x)\bigr]$.
Because $Y_n$ satisfies the equation $in\cdot D_s Y_n = 0$, the collinear quark Lagrangian in terms of the redefined fields $\xi_n^{(0)}$ contains no couplings to soft gluons whatsoever: $\bar\xi_n (in\cdot D_s) \xi_n \to \bar\xi_n^{(0}(in\cdot\partial)\xi_n^{(0)}$. 
A similar decoupling occurs in the collinear gluon Lagrangian. Thus SCET at leading order can be written in terms of entirely decoupled collinear and soft sectors. The interactions of collinear and soft fields in QCD appears in the form of operators containing both types of fields, as we will illustrate below, but they do not interact through the Lagrangian at all. It is this decoupling in the Lagrangian that makes soft-collinear factorization so simple to demonstrate in SCET at leading order in $\lambda$.

The theory \SCETa is appropriate for jets probed by their invariant mass $m^2\sim (Q\lambda)^2$ or similar measure (like thrust), which constrains soft radiation to have $p_s\sim Q\lambda^2$. When jets are instead probed by their internal transverse momentum (e.g. with jet broadening), the soft radiation is constrained to share the same $p_s^\perp\sim p_c^\perp\sim Q\lambda$ as collinear modes, giving them the same virtuality but different rapidity. \SCETb can be constructed by  matching QCD onto \SCETa and then lowering the virtuality of collinear modes to match the soft.\cite{Bauer:2002uv,Bauer:2003mga} In this theory, an additional regulator separating the soft and collinear modes in rapidity instead of virtuality is required in perturbative computations and RG running. The ``rapidity renormalization group'' is an elegant implementation of this idea\cite{Chiu:2011qc,Chiu:2012ir}, though not the only one\cite{Becher:2014oda,Becher:2010tm,GarciaEchevarria:2011rb,Echevarria:2012js}. To apply SCET to jets produced in heavy-ion collisions and propagating through a dense medium, e.g. quark-gluon plasma, a new mode needs to be added. Jet modification in a medium occurs through transverse kicks from scatterers in the medium through Glauber modes, of scale $Q(\lambda^2,\lambda^2,\lambda)$ in light-cone coordinates. The theory extended to include these modes is known as \SCETg\!.\cite{Idilbi:2008vm,DEramo:2010ak,Ovanesyan:2011xy}

\section{Factorization}
\label{sec:factorization}

To illustrate how factorization is carried out in general, we consider just one of many possible examples, jet production in deep inelastic scattering (DIS), $e(k)p(P) \to e(k')X$. We will consider measuring the final state with an event shape, DIS thrust\cite{Antonelli:1999kx}, or more generally, the ``1-jettiness,'' a special case of $N$-jettiness,\cite{Stewart:2010tn} which measures the degree to which the final state is collimated into $N$ distinct hadronic jets, plus beam radiation. Thus 1-jettiness in DIS measures collimation along the proton beam direction $q_B$ and another direction $q_J$:
\be
\label{eq:tau1}
\tau_1 = \frac{2}{Q^2} \sum_{i\in X} \min\{q_B\cdot p_i ,q_J\cdot p_i\}\,.
\ee
The min groups particles into two regions depending on which 4-vector $q_{B,J}$ they are closer to. The choice $q_B = xP$, $q_J = q+xP$, where $q=k-k'$, $x=-q^2/(2P\cdot q)$ yields $\tau_1 = \tau$, the classic DIS thrust.\cite{Antonelli:1999kx} We focus on this variable below. Other choices have also been studied\cite{Kang:2013nha,Kang:2013wca}. For small values $\tau_1\ll1$, the final state contains two well-collimated sets of particles in the beam and jet directions $q_B,q_J$. 

In QCD, the $\tau_1$ cross section can be expressed
\be
\label{eq:crosssection} 
\frac{d\sigma}{dx\,dQ^2\,d\tau_1} = L_{\mu\nu}(x,Q^2) W^{\mu\nu}(x,Q^2,\tau_1)\,,
\ee
where $L_{\mu\nu}$ is a leptonic tensor, and $W^{\mu\nu}$ is the hadronic tensor,
\be
\label{eq:Wtensor}
W_{\mu\nu}(x,Q^2,\tau_1) = \sum_X\bra{P} J_\mu^\dag\ket{X}\bra{X}J_\nu\ket{P}(2\pi)^4\delta^4(P+q-p_X)\delta(\tau_1-\tau_1(X))\,,
\ee
which is the usual hadronic tensor in DIS with an additional constraint on the 1-jettiness $\tau_1$ of the final state $X$. We will factor $W_{\mu\nu}$ in SCET into hard, collinear and soft contributions. This will then allow us to resum logs of $\tau_1$ in the cross section.  

We first match the QCD e.m. current $J^\mu = \bar q\gamma^\mu q$ onto operators in SCET. The matching condition takes the form
\be
\label{eq:matching}
J^\mu(x) = \! \sum_{n_1n_2} \!\int d^3\tilde p_1 d^3 \tilde p_2 e^{i(\tilde p_1 - \tilde p_2)\cdot x} C^\mu_{\alpha\beta}(\tilde p_1,\tilde p_2)\bar \chi^\alpha_{n_1,\tilde p_1}(x) T[Y_{n_1}^\dag Y_{n_2}](x)\chi^\beta_{n_2,\tilde p_2}(x)\,,
\ee
where $\chi_{n,\tilde p} = [ \delta(n\mcdot \tilde p \minus \bar n\mcdot\cP )\delta^2(\tilde p_\perp \minus \cP_\perp)W_n^\dag\xi_n]$ is a jet field with total label momentum equal to $\tilde p$, and where the soft fields Wilson lines $Y_{n_1,n_2}$ appear because we have already redefined the collinear fields according to \eq{BPS} (but omitted the $^{(0)}$ superscripts). In the sum over directions, the terms $n_{1,2}=n_{J,B}$ are selected out, with one line ($Y_{n_B}$) on an incoming path and the other ($Y_{n_J}$) outgoing for DIS. There are also sums over labels and spinor indices. The hard coefficients $C_{\alpha\beta}^\mu$ are determined order-by-order in $\as$ by requiring that matrix elements of the operators on the two sides be equal. (In general a color singlet gluon operator can also appear on the right-hand side, but we omit it here for simplicity.\cite{Kang:2013nha,Stewart:2009yx})

Computing matrix elements of the QCD operator on the left-hand side of \eq{matching} and of the SCET operator on the right-hand side with the same external states, one finds that the infrared behavior of the two sides is the same, while they differ in the ultraviolet. Requiring matrix elements of the two sides of \eq{matching} to be equal determines the value of the matching coefficient. To $\cO(\as)$,\cite{Bauer:2003di,Manohar:2003vb}
\be
\label{eq:coefficient}
C^\mu(\tilde p_1,\tilde p_2) = \gamma_\perp^\mu \biggl\{ 1 + \frac{\as(\mu)C_F}{4\pi}\biggl[ -\ln^2\Bigl(\frac{\mu^2}{-\tilde p_1\cdot \tilde p_2}\Bigr) - 3\ln\Bigl(\frac{\mu^2}{-\tilde p_1\cdot\tilde p_2}\Bigr)  - 8 + \frac{\pi^2}{6}\biggr]\biggr\}
\ee
Now, we substitute \eq{matching} into the hadronic tensor \eq{Wtensor}. In SCET, collinear fields in different directions and soft fields are all decoupled from one another in the leading-order Lagrangian after the field redefinition \eq{BPS}. Thus we obtain a factorized prediction for $W^{\mu\nu}$ in \eq{Wtensor} and thus for the cross section \eq{crosssection}:
\be
\label{eq:factorized}
\begin{split}
\frac{d\sigma}{dx\,dQ^2\,d\tau_1} &= \frac{d\sigma_0}{dx \,dQ^2}H_2(Q^2,\mu)\int dt_J\,dt_B\,dk_s\,\delta\Bigl(\tau_1 - \frac{t_J}{Q^2} - \frac{t_B}{Q^2} - \frac{k_s}{Q}\Bigr)  \\
&\quad \times  \int d^2\vect{p}_\perp S_{\text{hemi}}(k_s,\mu) J_q(t_J-\vect{p}_\perp^2,\mu) \cB_q(t_B,x,\vect{p}_\perp^2,\mu)\,,
\end{split}
\ee 
with an implicit sum over (anti)quark flavors $q$, and where $d\sigma_0/dx\,dQ^2$ the Born-level cross section. The hard function $H_2 = \lvert C(q^2)\rvert^2$ is given by the squared amplitude of the scalar part of the matching coefficient \eq{coefficient} for $\tilde p_1\mcdot\tilde p_2 = q^2=-Q^2$. The jet function $J_q$ is given by matrix element of jet fields:
\be
\begin{split}
J_q(t_J = \omega k^+  &\plus \omega_\perp^2,\mu) = \frac{(2\pi)^2}{N_C} \int \frac{dy^-}{2\abs{\omega}} e^{ik^+ y^-/2} \\
&\quad\times\tr \Bigl\langle{0}\Bigr\rvert \frac{\bn}{2} \chi_n(y^- n/2) \delta(\omega + \bn\cdot\cP)  \delta^2(\omega_\perp+\cP_\perp)\chi_n(0)\Bigl\lvert 0 \Bigr\rangle\,,
\end{split}
\ee
where $n$ is in the direction of $q_J$ and the trace is over Dirac indices. The matrix elements can be evaluated explicitly using Feynman rules of SCET. To $\cO(\as)$, the jet function is given by\cite{Bauer:2003pi}
\be
\label{eq:jetoneloop}
J_q(t,\mu) = \delta(t) + \frac{\as(\mu)}{4\pi} \Bigl\{(7-\pi^2)\delta(t) - \frac{3}{\mu^2}\Bigl[ \frac{\theta(t)}{t/\mu^2}\Bigr]_+ + \frac{4}{\mu^2}\Bigl[ \frac{\theta(t) \ln(t/\mu^2)}{t/\mu^2}\Bigr]_+\biggr\}\,,
\ee
where $[f(x)]_+$ is a plus distribution\cite{Ligeti:2008ac}. $J_q$ is known to $\cO(\as^2)$ and its anomalous dimension to $\cO(\as^3)$\cite{Becher:2006qw,Becher:2006mr} and contains logs of $t/\mu^2$. Meanwhile, $\cB_q$ is a beam function, given by the proton matrix element of collinear fields in the direction $n_B$ of $q_B$,
\be
\begin{split}
\cB_q\Bigl(\omega k^+,\frac{\omega}{P^-},\vect{k}_\perp^2,\mu\Bigr) &= \frac{\theta(\omega)}{\omega}\int\frac{dy^-}{4\pi} e^{ik^+ y^-/2} \bra{P} \bar\chi_n(y^- n_B/2)\frac{\bn_B}{2} \\
&\qquad\qquad  \times \Bigl[ \delta(\omega - \bn_B\cdot\cP) \frac{1}{\pi}\delta(k_\perp^2 - \cP_\perp^2) \chi_n(0)\Bigr]\ket{P}\,,
\end{split}
\ee
where the proton $P$ has momentum $P^- n_B/2$ and $y^-=n_B\mcdot y,k^+ = n_B\mcdot k$. The beam function can be matched onto ordinary parton distribution functions (PDFs) $f_j$, 
\be
\label{eq:beam}
\cB_q(t,x,\vect{k}_\perp^2,\mu) = \sum_j \int_x^1 \frac{d\xi}{\xi} \cI_{qj}\Bigl(t,\frac{x}{\xi},\vect{k}_\perp^2,\mu\Bigr) f_j(\xi,\mu) \Bigl[ 1 + \cO\Bigl( \frac{\Lqcd^2}{t},\frac{\Lqcd^2}{\vect{k}_\perp^2}\Bigr)\Bigr]\,,
\ee
with a sum over partons $j$ and where the coefficients $\cI_{qj}$ can be computed perturbatively\cite{Stewart:2009yx,Mantry:2009qz,Stewart:2010qs,Jain:2011iu,Gaunt:2014cfa,Gaunt:2014xga,Gaunt:2014xxa} and like the jet function in \eq{jetoneloop} contain logs of $t/\mu^2$. The soft function $S_{\text{hemi}}$ is given by matrix elements of soft Wilson lines,
\be
\label{eq:hemisoft}
\begin{split}
S^{ep}_{\text{hemi}}(k_s,\mu) &= \frac{1}{N_C}\Tr \sum_{X_s} \abs{\bra{X_s}T[Y_{n}^\dag Y_\bn](0) \ket{0}}^2  \delta \bigl( k_s - \bn\cdot k_X^L - n\cdot k_X^R  \bigr)\,,
\end{split}
\ee
where $k_X^{L,R}$ is the total momentum in the $\mp\vect{z}$ hemisphere of the final state $X$. The soft functions for two-jet production in $e^+e^-$ or $pp$ collisions can be obtained by turning the appropriate Wilson line paths to be incoming or outgoing.\cite{Arnesen:2005nk,Chay:2004zn} The $e^+e^-$ hemisphere soft function is known to $\cO(\as^2)$\cite{Fleming:2007xt,Kelley:2011ng,Monni:2011gb,Hornig:2011iu}. $S_{\text{hemi}}^{ee,ep,pp}$ have the same anomalous dimension to all orders. Although the two regions into which $\tau_1$ organizes particles may not be exact hemispheres, rescalings of $n_{J,B}$ and boost invariance allow use of the hemisphere soft function in \eq{hemisoft}.\cite{Kang:2013nha,Feige:2012vc} The soft function can be expressed as a convolution of a perturbative coefficient and a nonperturbative shape function, which captures the effects of hadronization in the final state\cite{Hoang:2007vb}:
\be
\label{eq:SNP}
S_{\text{hemi}}(k,\mu) = \int dk' S_{\text{hemi}}^{\text{pert}}(k-k',\mu) F(k')\,,
\ee
where $F$ is a nonperturbative shape function to be extracted from experiment.\cite{Abbate:2010xh,Ligeti:2008ac} The perturbative soft function contains logs of $k/\mu$.

Similar factorization theorems hold for other processes such as thrust or other event shape distributions in $e^+e^-$ or $pp$ collisions, \cite{Stewart:2009yx,Fleming:2007xt,Fleming:2007qr,Bauer:2008dt}
\be
\label{eq:thrustee}
\frac{1}{\sigma_0}\frac{d\sigma}{d\tau_{ee}} = H_2(-Q^2) J_n\otimes J_\bn \otimes S_{\text{hemi}}^{ee}\,, \quad \frac{1}{\sigma_0}\frac{d\sigma(q^2,Y)}{d\tau_B} = H_{ij}(q^2) B_i\otimes B_j \otimes S_{\text{hemi}}^{pp}
\ee
for $\tau_{ee} = 1-T$ where $T$ is the $e^+e^-$ thrust\cite{Farhi:1977sg} at collision energy $Q$, and $\tau_B$ is the beam thrust in Drell-Yan at dilepton invariant mass $q^2$ and rapidity $Y$.\cite{Stewart:2010tn,Stewart:2009yx,Stewart:2010pd} The $\otimes$ indicate all appropriate convolutions. These are similar to the DIS thrust factorization \eq{factorized}, with beam or jet functions appearing as appropriate for incoming or outgoing collinear modes. For $pp$ there is a sum over partonic channels $ij$. The soft functions are defined similarly to \eq{hemisoft} but in terms of incoming or outgoing Wilson lines, as appropriate. 
Thus cross sections for processes containing collinear jet or beam radiation in $ee,ep,pp$ collisions can all be written in terms of a simple set of building blocks in SCET.

\section{Evolution and Resummation}
\label{sec:resummation}

The factorization of a multi-scale cross section like \eq{factorized} allows the resummation of large logarithms of ratios of these scales that appear in the fixed-order perturbative expansion of the cross section in full QCD. After performing the convolutions in \eq{factorized}, one finds that the hard function contributes logs of $\mu/Q$, the jet and beam functions logs of $\mu/(Q\tau^{1/2})$, and the soft function logs of $\mu/(Q\tau)$. The dependence on $\mu$ cancels in the full cross section, but logs of $\tau$ are left over. 
These logs in the fixed-order expansion of cross sections \eqs{factorized}{thrustee} take the generic form\cite{Catani:1992ua}
\be
\label{eq:fixedorder}
\begin{split}
\ln\sigma(\tau) = \quad  \frac{\as}{4\pi} \quad \Bigl( F_{12}L_\tau^2 &+ F_{11}L_\tau  + F_{10}\Bigr) \\
+\Bigl(\frac{\as}{4\pi}\Bigr)^2 \Bigl(F_{23} L_\tau^3 &+ F_{22}L_\tau^2 + F_{21} L_\tau + F_{20} \Bigr) \\
+\Bigl(\frac{\as}{4\pi}\Bigr)^3 \Bigl( F_{34} L_\tau^4 &+ F_{33}L_\tau^3 + F_{32} L_\tau^2 + F_{31}L_\tau + F_{30}\Bigr)+ \cdots \,, \\
\text{\small LL} \ &  \qquad \text{\small NLL} \quad \  \text{\small NNLL} \quad \   \text{N$^3$LL}
\end{split}
\ee
where $\sigma(\tau) \equiv (1/\sigma_0)\int_0^\tau d\tau' (d\sigma/d\tau')$, $L_\tau\equiv \ln(1/\tau)$, and where the dots indicate higher-order terms in $\as$ and non-singular terms which vanish as $\tau\to 0$. Fixed-order perturbation theory sums this series row-by-row in $\as$, which blows up for small $\tau$. In this regime, the appropriate organization is resummed perturbation theory, which sums up each whole column at a time. The first column is the set of \emph{leading logs} (LL), the second column the \emph{next-to-leading logs} (NLL), and so forth. 

The factorization theorems \eqs{factorized}{thrustee} offer a systematic way to achieve this resummation. Taking the Laplace (alternatively, Fourier) transform of each in $\tau$, we undo the convolutions and obtain, in the case of $e^+e^-$ thrust,
\be
\wt\sigma(\nu) = H(Q^2,\mu) \wt J_n\Bigl(\frac{\nu}{Q^2},\mu\Bigr) \wt J_\bn\Bigl(\frac{\nu}{Q^2},\mu\Bigr) \wt S_{\text{hemi}}^{ee}\Bigl(\frac{\nu}{Q},\mu\Bigr)\,,
\ee
where the Laplace transforms are given by $\tilde f(\nu/Q^j) = \int_0^\infty dx \,e^{-x\nu/Q^j} f(x)$ for each function $\sigma,J,S$ with argument $x=\tau,t_{n,\bn},k_s$ with mass dimension $j = 0,2,1$, respectively. The fixed-order expansions of these functions contain logs of $\nu/Q^j$. They each obey an RG evolution equation,
\be
\label{eq:RGE}
\frac{d}{d\ln\mu} \tilde f(\nu,\mu) = \wt\gamma_f(\mu) \tilde f(\nu,\mu)\,,
\ee
where $\tilde \gamma_f(\mu)$ is anomalous dimension of $\tilde f$, which takes the form
\be
\label{eq:anomdim}
\wt\gamma_f(\nu,\mu) = -\kappa_f \Gamma_{\text{cusp}}[\as] \ln(\mu^{j_f} \nu e^{\gamma_E}) + \gamma_f[\as]\,, 
\ee
where $\Gamma_{\text{cusp}}[\as]$ is the cusp anomalous dimension, $\gamma_f[\as]$ is the non-cusp part of the anomalous dimension, and $\kappa _S= 1$ and $\kappa_{J,B} = 2$. The hard function obeys a similar equation to \eq{RGE} with $j_H = 1$ and $\nu e^{\gamma_E}\to 1/Q$. 
The RGE \eq{RGE} is easily solved, and gives the value of $\tilde f$ at one scale $\mu$ in terms of $\tilde f$ at another scale $\mu_0$,
\be
\label{eq:RGEsol}
\tilde f(\nu,\mu) = \tilde f(\nu,\mu_0)\exp\biggl[\int_{\mu_0}^\mu \frac{d\mu'}{\mu'}  \wt\gamma_f(\nu,\mu')\biggr]\,.
\ee
A more explicit version of the integral in the exponent, and the inverse transforms back to momentum space, can easily be found in the literature, e.g. [\refcite{Almeida:2014uva}]. In momentum space, these RGEs lead to the prediction for, e.g. the $e^+e^-$ thrust distribution,
\begin{align}
&\sigma(\tau) = H_2(Q^2,\mu_H) U_H(\mu,\mu_H)\! \int \! dt_n dt_n' dt_\bn dt_\bn' dk_s dk_s' \delta\Bigl( \tau - \frac{t_n+t_\bn}{Q^2} - \frac{k_s}{Q}\Bigr) S(k_s',\mu_S) \nn \\
& \times \! J_n(t_n',\mu_J)U_{J_n}(t_n \minus t_n',\mu,\mu_J)J_\bn(t_\bn',\mu_J) U_{J_\bn}(t_\bn \minus t_\bn',\mu,\mu_J)U_S(k_s\minus k_s',\mu,\mu_S),
\end{align}
where the evolution kernels $U_f$ from $\mu_f$ to $\mu$ are given by solutions \eq{RGEsol} of the RGE, transformed back to momentum space. This form allows us to evaluate the individual functions $H,J_{n,\bn},S$ at scales $\mu_f$ where logs in their fixed-order expansions are small, and then the evolution kernels exponentiate the infinite sets of large logs in \eq{fixedorder} in a closed form. The accuracy to which the logs are summed is determined by the accuracy to which the anomalous dimensions in \eq{anomdim}, the beta function for running $\as$, and factorization ingredients $f=H,J,B,S$ are known. Namely, at N$^k$LL accuracy, $\Gamma_{\text{cusp}}$ is needed to $\cO(\as^{k+1})$, $\gamma_f$ to $\cO(\as^k)$, the beta function $\beta[\as]$ to $\cO(\as^{k+1})$, and the fixed-order constants in the factorization ingredients $f=H,J,B,S$ to $\cO(\as^{k-1})$. One can define the ``primed'' counting N$^k$LL$'$ by including $f$ to one higher order, $\as^k$, which offers better matching onto the large $\tau$ region where we revert to fixed-order perturbation theory,\cite{Abbate:2010xh} and helps maintain closer agreement in accuracy between momentum- and Laplace-space cross sections. \cite{Almeida:2014uva}

\section{Nonperturbative Effects}
\label{sec:nonperturbative}

The nonperturbative effects of hadronization on the thrust distributions in $ee,ep,pp$ in \eqs{factorized}{thrustee} are contained in the soft function, \eq{SNP}. The definition of the soft function in \eq{hemisoft} (and its analogs for $ee,pp$) show that it is independent of the hard scale $Q$ and depends only on the directions and color representations of the hard particles initiating the collinear jets.

In the ``tail region'' where $\tau \gg \Lqcd/Q$ but $\tau\ll 1$, an even stronger level of universality holds. In this region, the soft shape function in \eq{SNP} can be expanded in an operator product expansion, $F_e(k) = \delta (k) - \delta'(k) \Omega_1^e + \cdots$, where the dots indicate power corrections and perturbative corrections, and $e$ specifies the event shape. The leading effect is then a shift of the first moment of the whole distribution by an amount $\Omega_1^e/Q$. The parameter $\Omega_1^e$ is defined by
\be
\label{eq:ANP}
\Omega_1^e = \int_0^\infty d\eta f_e(\eta) \frac{1}{N_C}\Tr \bra{0}\overline T[ Y_\bn Y_n^\dag](0) \mathcal{E}_T(\eta) T [Y_n Y_\bn^\dag](0)\ket{0}\,,
\ee
where $\cE_T(\eta)$ is a ``transverse energy flow'' operator\cite{Bauer:2008dt,Sveshnikov:1995vi,Cherzor:1997ak,Belitsky:2001ij} for particles with (pseudo)rapidity $\eta$ with respect to $n$, and $f_e$ is a weighting function whose definition depends on the event shape $e$. For thrust $\tau$, $f_\tau= 2e^{-\abs{\eta}}$. This form assumes massless hadrons in the final state; the appropriate form for massive hadrons involves the use of a transverse velocity operator\cite{Mateu:2012nk}. In the form \eq{ANP}, $\Omega_1^e$ appears to depend nontrivially on the event shape $e$. However, the matrix element itself is independent of $\eta$ thanks to boost invariance of the Wilson lines $Y_{n,\bn}$.\cite{Lee:2006fn,Lee:2006nr} This means $\Omega_1^e$ takes the form $\Omega_1^e = c_e \Omega_1$, where $c_e = \int_0^\infty d\eta\, f_e(\eta)$ and $\Omega_1= \frac{1}{N_C}\Tr \bra{0}\overline T[ Y_\bn Y_n^\dag](0) \mathcal{E}_T(0) T [Y_n Y_\bn^\dag](0)\ket{0}$. So $c_e$ is just a number, exactly calculable, and the nonperturbative parameter $\Omega_1$ itself is universal, not only in $Q$ but for multiple event shapes $e$ in a given process, such as angularities in $e^+e^-$ \cite{Lee:2006nr,Berger:2003iw,Berger:2003pk,Hornig:2009vb}, or different versions of 1-jettiness in DIS.\cite{Kang:2013nha} The universality does not necessarily hold upon changing the directions of the lines $Y_{n,\bn}$ and thus not across $ee,ep,pp$. However, a remarkable relation was discovered between the $\Omega_1$ for DIS 1-jettiness and the leading soft power correction to jet mass for small $R$ in $pp\to Z/H+\text{jet}$.\cite{Stewart:2014nna}

Thus SCET provides a powerful framework to analyze rigorously properties of nonperturbative effects. The proof of the universality of $\Omega_1$ in $e^+e^-$ event shapes, for instance, generalizes earlier arguments based on models of an ``effective infrared coupling'' at low scales and the emission of only a finite number of soft gluons.\cite{Dokshitzer:1997ew,Dokshitzer:1998pt}

\section{Recent progress and future directions}
\label{sec:recent}

Here we mainly reviewed the basic elements of \SCETa and its primary applications to the resummation of large logarithms in QCD perturbation theory and the analysis of nonperturbative corrections. We have almost entirely overlooked  the treatment of jet algorithms in SCET,\cite{Ellis:2010rw} and applications of \SCETb and \SCETg\!\!, such as resummation of jet broadening\cite{Chiu:2012ir,Becher:2011pf}, transverse momentum dependent parton distributions\cite{GarciaEchevarria:2011rb,Echevarria:2012js}, jet modification in dense media\cite{Idilbi:2008vm,DEramo:2010ak,Ovanesyan:2011xy,Chien:2014nsa,Vitev:2014kha,Chien:2014zna}, as well as many others we are regrettably unable to cite here. Ref.~[\refcite{Becher:2014oda}] has a fairly comprehensive list.

Glauber modes also enter the analysis of factorization theorems in the vacuum, and their cancellation in virtual loops is essential to factorizing collinear modes in different directions.\cite{Collins:1981ta,Collins:1983ju,Collins:1985ue,Collins:1988ig,Bauer:2010cc} Recently progress has been made incorporating these modes into SCET analyses of hard scattering processes in the vacuum and their connection to Regge behavior and the BFKL equation.\cite{Donoghue:2009cq,Donoghue:2014mpa,Fleming:2014rea,Gaunt:2014ska}

SCET has proven to be a revolutionary tool in the analysis of hard scattering cross sections, in particular those containing jets or collinear beam radiation. The number of applications in high-energy and nuclear physics is ever growing. As we continue to use jets as probes of QCD and new physics at LHC, RHIC, and an EIC, SCET in its various flavors will continue to prove instrumental in our improved theoretical understanding of both perturbative and nonperturbative physics.

\section*{Acknowledgments}

I would like to thank my collaborators, especially Christian Bauer, Steve Ellis, Sean Fleming, Andrew Hornig, Daekyoung Kang, Aneesh Manohar, Greg Ovanesyan, George Sterman, Iain Stewart, Jon Walsh, and Mark Wise, whose work is highlighted above and who have been important influences on my knowledge and perspective about EFT, QCD, and SCET. I thank I. Stewart, the originally scheduled speaker, for providing some material for this talk. This work is supported in part by DOE Contract DE-AC52-06NA25396 and by the LDRD office at Los Alamos.

\bibliographystyle{ws-ijmpcs} 
\bibliography{scet}

\end{document}